\documentclass[fleqn,usenatbib]{mnras}

\usepackage[T1]{fontenc}

\DeclareRobustCommand{\VAN}[3]{#2}
\let\VANthebibliography\thebibliography
\def\thebibliography{\DeclareRobustCommand{\VAN}[3]{##3}\VANthebibliography}

\usepackage{graphicx}	
\usepackage{amsmath}	
\usepackage{amssymb}	
\usepackage{animate}

\newcommand{\Msun}{M$_{\odot}$\,}
\newcommand{\Mstar}{M$_{*}$\,}
\newcommand{\BtoTkin}{(B/T)$_{\text{kin}}$\,}

\newcommand{\BtoTkinOneGyrBefore}{(B/T)$_{\text{kin,-1Gyr}}$\,}
\newcommand{\BtoTkinOneGyrAfter}{(B/T)$_{\text{kin,+1Gyr}}$\,}
\newcommand{\rd}{revived disc}
\newcommand{\sd}{survived disc}
\newcommand{\cd}{constant disc}
\newcommand{\rds}{revived discs}
\newcommand{\sds}{survived discs}
\newcommand{\cds}{constant discs}
\newcommand{\Rd}{Revived disc}

\title[Massive disc galaxies formation]{Formation of massive disc galaxies in the IllustrisTNG simulation}

\author[Zeng et al.]{
Guangquan Zeng,$^{1,2}$\thanks{E-mail: zenggq@nao.cas.cn}
Lan Wang,$^{1}$\thanks{E-mail: wanglan@bao.ac.cn}
Liang Gao$^{1,2,3}$
\\
$^{1}$Key Laboratory for Computational Astrophysics, National Astronomical Observatories, Chinese Academy of Sciences, Beijing 100101, China \\
$^{2}$School of Astronomy and Space Science, University of Chinese Academy of Sciences, Beijing 100049, China \\
$^{3}$Institute of Computational Cosmology, Department of Physics, University of Durham, South Road, Durham DH1 3LE, UK \\
}

\date{Accepted XXX. Received YYY; in original form ZZZ}

\pubyear{2021}

\begin{document}
\label{firstpage}
\pagerange{\pageref{firstpage}--\pageref{lastpage}}
\maketitle

\begin{abstract}
We investigate the formation history of massive disc galaxies in hydrodynamical simulation -- the IllustrisTNG, to study why massive disc galaxies survive through cosmic time. 83 galaxies in the simulation are selected with M$_{*,z=0}$ $>8\times10^{10}$ \Msun and kinematic bulge-to-total ratio less than $0.3$. We find that 8.4 percent of these massive disc galaxies have quiet merger histories and preserve disc morphology since formed. 54.2 percent have a significant increase in bulge components in history, then become discs again till present time. The rest 37.3 percent experience prominent mergers but survive to remain discy. While mergers and even major mergers do not always turn disc galaxies into ellipticals, we study the relations between various properties of mergers and the morphology of merger remnants. We find a strong dependence of remnant morphology on the orbit type of major mergers. Specifically, major mergers with a spiral-in falling orbit mostly lead to disc-dominant remnants, and major mergers of head-on galaxy-galaxy collision mostly form ellipticals. This dependence of remnant morphology on orbit type is much stronger than the dependence on cold gas fraction or orbital configuration of merger system as previously studied.
\end{abstract}



\section{Introduction}
\label{sec:Intro}

Since \citet{hubble1926extragalactic} proposed the classification scheme on galaxy morphology, extensive studies have been performed to understand how galaxies evolve into different morphological states in the Universe today. In general, disc galaxies are thought to obtain their original angular momentum through tidal torques in the early Universe and form the disc structures by the dissipational collapse of gas in their host dark matter haloes \citep[][]{peebles1969origin, doroshkevich1970spatial, fall1979dissipation, fall1980formation, mo1998formation}, and elliptical galaxies are considered as the products of mergers between galaxies \citep[][]{toomre1977theories, white1978core, fall1979dissipation}. Under this basic scenario, massive galaxies at low redshift are likely to have an elliptical morphology, since they should have experienced many mergers as predicted by the standard hierarchical model of galaxies formation. This prediction is statistically consistent with observations \citep[e.g.][]{buitrago2013early, conselice2014evolution, van20143d}. Nevertheless, a number of massive galaxies with a disc-like morphology at low redshift were also reported in observations \citep[e.g.][]{ogle2016superluminous, ogle2019catalog, luo2020has}.

Some of the massive disc galaxies are able to maintain a discy structure because they have quiet merger history, without experiencing prominent mergers in their lifetimes. 
For example, 
\citet{font2017diversity} studied disc galaxies with stellar mass around $10^{10}$ \Msun in the GIMIC simulations and found that some of the galaxies have no mergers since $z=2$. 
\citet{jackson2020extremely} analysed disc galaxies with stellar mass greater than $10^{11.4}$ \Msun in the Horizon-AGN simulation, and found that 30 per cent of them are always disc due to quiet merger history. 
Other massive disc galaxies, however, have inevitably experienced times of mergers but somehow do not transform to early-type galaxies at present day.

A lot of studies based on numerical simulations have shown great complexity between merger properties and the morphological transformation. Although merger is thought to take responsibility for the transformation from disc into spheroid, it is found that even major merger does not necessarily convert a disc galaxy into a bulge-dominant one \citep[e.g.][]{springel2005formation, hopkins2009disks, governato2009forming, athanassoula2016forming, eliche2018formation, wang2019comparing}. In addition to the mass ratio of the merging pair, morphology of the remnant is related also to the (cold) gas fraction, the orbital configuration of the merging system (e.g. prograde versus retrograde), and the original morphology of progenitor galaxies, etc \citep[e.g.][]{barnes1996transformations, naab2003statistical, cox2006kinematic, robertson2006merger, hopkins2009disks}. A gas-rich merger leads to a disc-like remnant more frequently than a gas-poor merger \citep[e.g.][]{lotz2008galaxy, font2017diversity, martin2018role, garrison2018origin, peschken2020disc}. Mergers with prograde orbital configuration are more likely to preserve the original disc morphology \citep[e.g.][]{hopkins2009disks, martin2018role, saburova2018malin, saburova2021observational}. Besides, \citet[][]{sparre2017unorthodox} showed that when AGN feedback is not strong enough, the products of major mergers may evolve more into star-forming disc galaxies.

An effect of merger property on morphology transformation that has not been considered statistically is the type of merger orbit, in the sense that whether the two galaxies have head-on collision, or one galaxy spirals in around the other gradually. Merger orbit may have an important effect on the final morphology of the merger remnant. Head-on collisions can cause disruption of the pre-existing discs and lead to bulge-dominant remnants. On the other hand, spiral-in falling is less violent and may maintain the original disc structure more often. This effect is however, not considered in detail in previous works which mainly focused on the effect of cold gas fraction, orbital configuration, etc. Therefore in this work, when studying the morphology evolution of massive disc galaxies, we pay special attention to the effect of merger orbit type on the morphology of post-merger galaxies.

In this work, we use the updated version of Illustris, the hydro-dynamical simulation IllustrisTNG \citep[][]{marinacci2018first, pillepich2018first, naiman2018first, springel2018first, nelson2018first} to explore the morphology evolution of massive disc galaxies, and compare their morphology evolution and merger history with massive elliptical galaxies. For massive galaxies that have experienced mergers, especially recent major mergers, we study in detail the relation between different merger properties and the morphology change of galaxies caused by merger, to investigate why sometimes mergers result in bulge-dominant remnants and sometimes not. 

This paper is organized as follows. In Section \ref{sec:Sample}, we introduce the IllustrisTNG simulation used in this work, how we define major and minor mergers of galaxies based on the simulation subhalo merger trees, and how we select massive galaxies with different morphologies. In Section \ref{sec:MorphologyEvolution}, we compare morphology evolution across cosmic time for different morphology types of massive galaxies selected at $z=0$. In Section \ref{sec:MorphologyChange}, we explore the morphology change of massive galaxies before and after their latest major mergers, and study the dependence of morphology change on cold gas fraction, orbital configuration, and in particular the orbit type of mergers. Finally, discussion and conclusions are presented in Section \ref{sec:Conclusions}.

\section{Simulation and Sample selection}
\label{sec:Sample}

\subsection{TNG100-1 simulation}
\label{subsec:TNG100-1}

The IllustrisTNG project \citep[][]{springel2018first, nelson2018first, naiman2018first, marinacci2018first, pillepich2018first}, which is the successor of the Illustris project \citep[][]{vogelsberger2014properties, vogelsberger2014introducing, genel2014introducing, sijacki2015illustris}, is a suite of cosmological magnetohydrodynamical simulations of galaxy formation, run with the moving-mesh code \texttt{AREPO} \citep[][]{springel2010pur}. IllustrisTNG is able to reproduce the statistical properties of galaxy morphology in good agreement with observations \citep[e.g.][]{tacchella2019morphology}. Besides, the size evolution of galaxies simulated by IllustrisTNG is more realistic than the original Illustris simulation \citep[e.g.][]{genel2018size}.

All the simulation data from the IllustrisTNG project have been released \citep[][]{nelson2019illustristng}. In this work, we use the publicly available results from TNG100-1, which simulates a periodic cubic volume with a side length of 110.7 Mpc, with both initial number of gas cells and dark matter particles to be $1820^3$, implementing the fiducial TNG physics model \citep[][]{weinberger2016simulating, pillepich2018simulating}. In TNG100-1, the mass resolution of dark matter particle is $7.5\times10^{6}$ \Msun, and the initial baryonic mass resolution is  $1.4\times10^{6}$ \Msun.

The \texttt{FoF} and \texttt{Subfind} algorithms \citep[][]{springel2005simulations, dolag2009substructures} are used to identify haloes and subhaloes for each snapshot. Merger trees are constructed by the \texttt{SubLink} algorithms \citep[][]{rodriguez2015merger}, to trace the progenitor(s) and a unique descendant of each subhalo. The main progenitor of a subhalo is defined as its most massive progenitor, and by tracing main progenitors in different snapshots back in time, the main progenitor branch of a given subhalo can be obtained. Simulated galaxies are associated with subhaloes, and their formation histories can be derived based on the merger trees of subhaloes.

If two galaxies share the same descendant, a merger between these two is considered to happen.
In the process of two galaxies approaching to each other, the stellar mass of the secondary galaxy can decrease by quite an amount before they finally merge \citep[e.g.][]{wang2019comparing, peschken2020disc}.
This is because that as the satellite gradually approaches to the central galaxy, the material of the satellite can be stripped and may be partially accreted by the central. Therefore, to get a mass ratio of the merging galaxy pair that reflects the relative size of galaxies when they are well isolated to each other before interaction starts, for each galaxy, we trace back the main progenitor branch and use the maximum stellar mass\footnote{In this work, stellar mass of a galaxy is defined as the total mass of stellar particles contained within twice the stellar half-mass radius of the galaxy.} in their histories to do the calculation. 
Throughout this work, major mergers are defined as mergers with stellar mass ratio of galaxies greater than $1:4$, and those with stellar mass ratio in the range of $1:10$ to $1:4$ are considered as minor mergers.

\subsection{Selection of massive disc galaxies}
\begin{figure*}
    \hspace{-0.4cm}
    \resizebox{17cm}{!}{\includegraphics{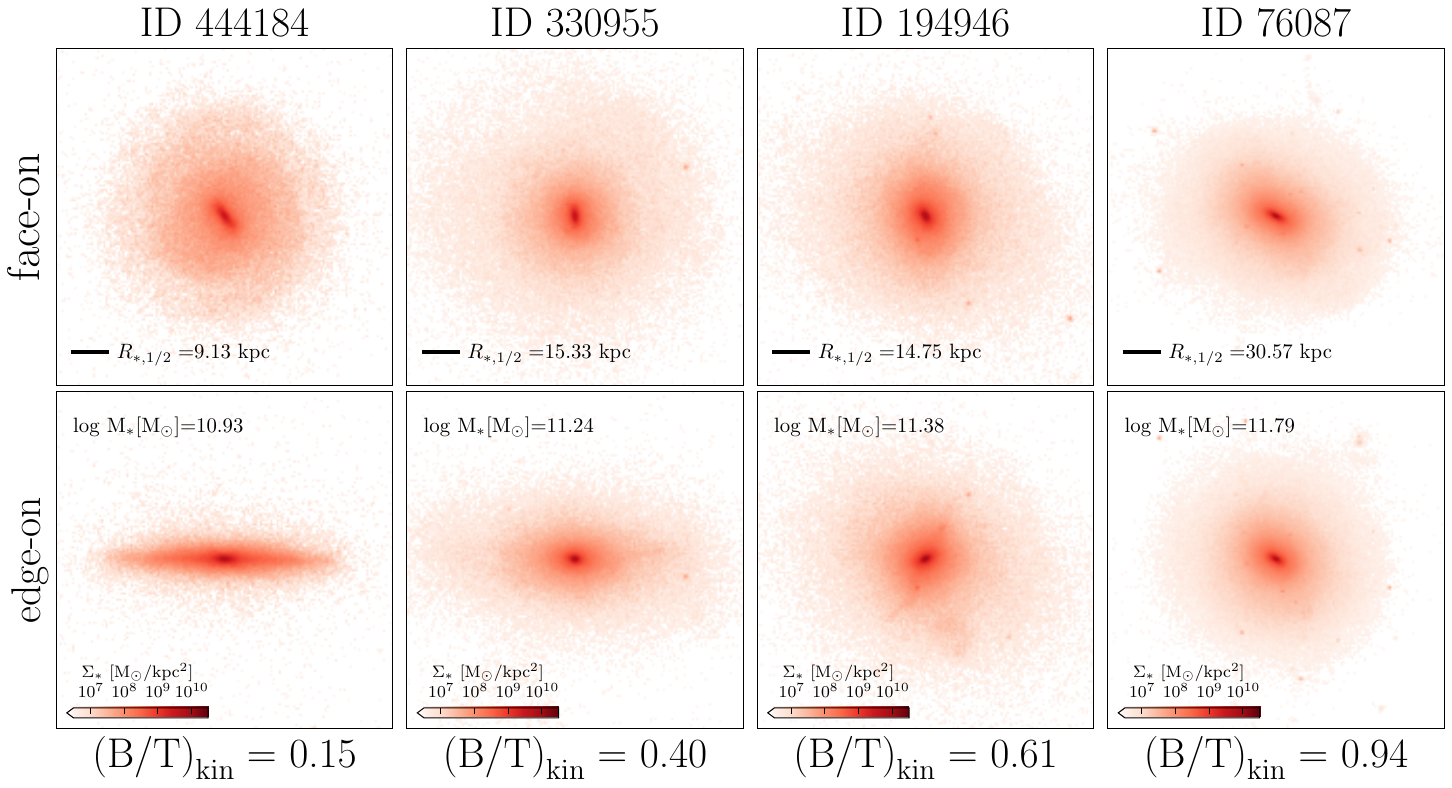}}
        \caption{Stellar mass density projections of four example galaxies with kinematic bulge-to-total ratio of 0.15, 0.40, 0.61 and 0.94 at $z=0$ in the TNG100-1 simulation. The upper row shows the face-on projections of these galaxies, and the bottom row shows the edge-on projections. For each galaxy, the subfind ID and \BtoTkin are listed above and below each column. The stellar mass, half stellar mass radius and scale of the stellar mass surface density of each galaxy are indicated in the corresponding panels. From left to right, it can be seen that as \BtoTkin increases, galaxy morphology changes from disc to spheroidal.}
        \label{fig:from_disk_to_elliptical}
\end{figure*}

\begin{figure}
	\includegraphics[width=\columnwidth]{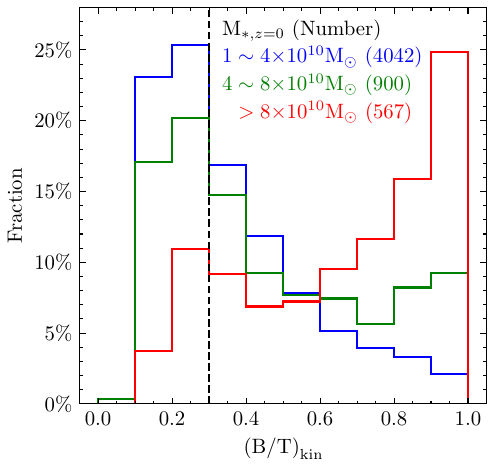}
    \caption{Distributions of kinematic bulge-to-total stellar mass ratio for galaxies at $z=0$ in TNG100-1, in three different stellar mass bins: $1 \sim 4 \times 10^{10}$ \Msun (blue), $4 \sim 8 \times 10^{10}$ \Msun (green) and $> 8 \times 10^{10}$ \Msun (red). The numbers of galaxies in each stellar mass bin are indicated in different colours accordingly. The black dashed vertical line indicates the value of \BtoTkin $=0.30$. Galaxies with \Mstar $>8\times10^{10}$ \Msun and \BtoTkin $< 0.3$ at $z=0$ are selected as massive disc galaxy sample. For comparison, galaxies with \Mstar $>8\times10^{10}$ \Msun and \BtoTkin $> 0.9$ at $z=0$ are selected as massive elliptical galaxies.}
    \label{fig:BtoT_kin_distribution_at_z0}
\end{figure}

We use a kinematics-based bulge-to-total stellar mass ratio to quantify morphology of simulated galaxies.
For a given stellar particle in each galaxy, the circularity parameter $\epsilon$ is defined as the ratio of its specific angular momentum to that of the circular orbit of the same radius. Then, bulge-to-total stellar mass ratio of a galaxy is defined by twice the mass fraction of stars with $\epsilon<0$ measured within 10 times the effective radius of the galaxy.
One can refer to the website of the IllustrisTNG project\footnote{\url{https://www.tng-project.org/}} and \citet{genel2015galactic} for more detailed description. 
In Fig.~\ref{fig:from_disk_to_elliptical}, we give a few examples of galaxies with different bulge-to-total ratio defined as described above. With the ratio increasing from left to right, the visual morphology of galaxies changes from disc to spheroidal shape. 

Statistically, morphological types of galaxies depends on their stellar mass.
For 5509 galaxies with stellar mass greater than $10^{10}$ \Msun at $z=0$ in TNG100-1 simulation, we divided them into three different stellar mass bins: $1 \sim 4\times10^{10}$ \Msun, $4 \sim 8\times10^{10}$ \Msun and $>8\times10^{10}$ \Msun, which correspond to galaxies less massive, of similar mass, and more massive compared with the Milky Way, respectively. As shown in Fig.~\ref{fig:BtoT_kin_distribution_at_z0}, a larger fraction of less massive galaxies have small \BtoTkin. For galaxies more massive than $8\times10^{10}$ \Msun, the distribution of \BtoTkin peaks at $0.9 \sim 1.0$, and most of them are bulge-dominant early type or ellipticals. Nevertheless, there exists a small peak at around \BtoTkin$=0.3$ in the distribution. A fraction of massive galaxies in simulation have a discy morphology, consistent with that discovered in observation as mentioned in Section \ref{sec:Intro}.

In the following, we focus on the morphological evolution of massive galaxies with stellar mass M$_{*,z=0}$ $>8\times10^{10}$ \Msun, which amount to 567 galaxies in the TNG100-1. Massive galaxies with \BtoTkin $< 0.3$ are selected as our sample of massive disc galaxies. For comparison, massive elliptical galaxies are defined as those with \BtoTkin $> 0.9$. By doing so, we select out 83 massive disc galaxies and 142 massive elliptical galaxies at $z=0$ in the TNG100-1.

\section{Morphology evolution of massive disc galaxies}
\label{sec:MorphologyEvolution}

Based on the massive galaxy sample selected, we trace back their formation histories, and study their morphology evolution through cosmic time. In this section, we first compare the general trend of morphology evolution and merger histories of massive disc galaxies with other massive galaxies, then we look into details of different pathways of how massive disc galaxies evolve to their present morphology.

\subsection{Morphology evolution and merger history of massive galaxies}

\begin{figure*}
    \hspace{-0.4cm}
    \resizebox{17cm}{!}{\includegraphics{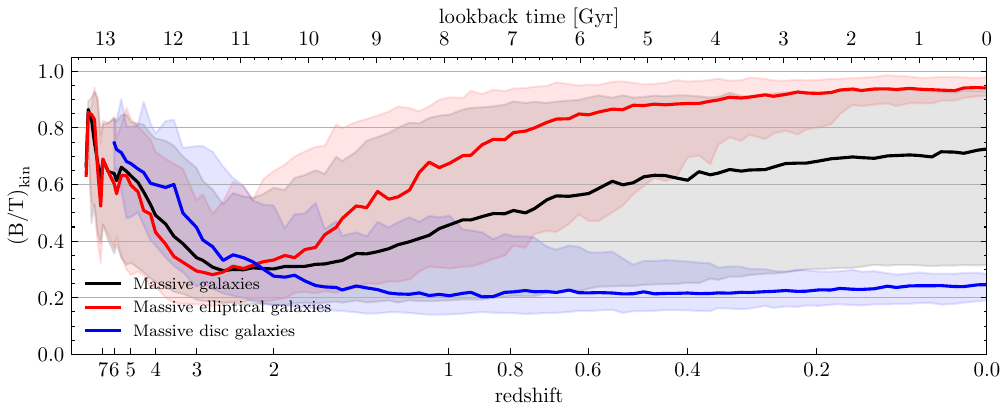}}
        \caption{The evolution of \BtoTkin for the main progenitors of all massive galaxies (black), massive elliptical galaxies (red) and massive disc galaxies (blue) selected at $z=0$ in TNG100-1 simulation. The solid lines of different colours show the median \BtoTkin for galaxies in each sample, and the shadow regions include the 16th-84th percentile distributions.}
        \label{fig:massive_galaxies_BtoT_kin_evolution}
\end{figure*}

For each of the massive galaxies in our sample, we trace back along its main progenitor branch, as described in Section \ref{subsec:TNG100-1}, and record the kinematic bulge-to-total ratio \BtoTkin for each main progenitor, to indicate morphology evolution of the galaxy through cosmic time. The results for all massive galaxies, massive disc galaxies, and massive elliptical galaxies are presented in Fig.~\ref{fig:massive_galaxies_BtoT_kin_evolution}. Solid lines give the median \BtoTkin of main progenitors for each sample of galaxies, and shadow regions include 16th-84th percentiles. A clear trend shown in Fig.~\ref{fig:massive_galaxies_BtoT_kin_evolution} is that, the progenitors of massive galaxies in general have large bulge fractions at high redshifts, and the bulge fractions decrease with time till redshift of around 2-3, regardless of their morphology at present day. A similar trend is also found in \citet[Figure 13]{correa2020dependence}.

\begin{table*}
\centering
\caption{For massive galaxies with different morphologies at present day, the average numbers of major (minor) mergers the galaxies experienced since formed, at $z \geq 2$, at $1 \leq z < 2$ and at $z < 1$ are listed. Massive disc galaxies are further divided into three subsamples, according to their different morphology evolution pathways (see Section \ref{subsec:DifferentPathways} for details). In the bottom row of the table, the numbers of galaxies in different samples are shown.}
\label{tab:merger_number}
\begin{tabular}{ccccccc}
\hline
\multicolumn{1}{l}{} & \multicolumn{6}{c}{Massive galaxies}                                                \\ \cline{2-7} 
                     & All          & Ellipticals  & \multicolumn{4}{c}{Discs}                             \\ \cline{4-7} 
                     &              &              & all         & \rds        & \sds        & \cds        \\ \hline
Since formed         & 3.16 (3.44)  & 3.55 (3.50)  & 2.61 (3.16) & 2.93 (3.47) & 2.32 (2.90) & 1.86 (2.29) \\ \hline
$z \geq 2$           & 1.78 (2.03)  & 1.81 (2.20)  & 1.75 (1.72) & 1.73 (1.73) & 1.77 (1.65) & 1.71 (2.00) \\ \hline
$1 \leq z < 2$       & 0.53 (0.54)  & 0.63 (0.45)  & 0.46 (0.55) & 0.60 (0.58) & 0.32 (0.58) & 0.14 (0.29) \\ \hline
$z < 1$              & 0.85 (0.86)  & 1.11 (0.85)  & 0.41 (0.88) & 0.60 (1.16) & 0.23 (0.68) & 0.00 (0.00) \\ \hline
\hline
Number of samples    & 567          & 142          & 83          & 45          & 31          & 7           \\ \hline
\end{tabular}
\end{table*}

After $z \sim 2$, for the overall sample of massive galaxies selected at $z=0$, the bulge fraction increases gradually. On average, these galaxies change from disc-dominated at $z\sim2$ to bulge-dominated with \BtoTkin$>0.6$ at $z<0.4$. The trend is in agreement with some observational results  \citep[e.g.][]{buitrago2013early}.
For massive galaxies with distinct morphologies at present day, they show quite different evolution of \BtoTkin. Specifically, massive ellipticals grow their bulge components relatively fast. At $z=1$, most of the massive ellipticals selected today have  \BtoTkin$>0.6$, already in a state of early type in morphology (see Fig.~\ref{fig:from_disk_to_elliptical} for reference). These galaxies gradually increase their bulge component till today. In contrast, the main progenitors of the massive disc galaxies keep decreasing their bulge fraction after $z \sim 2$, and most of them remain in a low bulge fraction state with \BtoTkin$<0.3$ at $z < 2$.

To explore what causes the distinct evolution of morphology for massive disc and massive elliptical galaxies after about redshift 2, we first check and compare the assembly and merger statistics of these galaxies. By tracing back the main progenitor branch of each massive galaxy in our sample, we record the merger events the galaxy experience in history. Based on the definition of major and minor mergers as described in Section \ref{subsec:TNG100-1}, we calculate the average number of both major and minor mergers that happened for all massive galaxies, massive discs, and massive ellipticals, respectively. The results are listed in Table~\ref{tab:merger_number} for mergers the galaxies experienced since they formed and in different redshift intervals, respectively.

On average, massive elliptical galaxies experienced $\sim 1$ more major merger than their disc counterparts in their lifetimes. The excess is contributed mainly by mergers after $z=1$. On the other hand, the average number of minor mergers experienced by massive discs is similar to that of massive ellipticals and of the whole sample. From these numbers, it seems that there is no strong correlation between minor merger history and the current kinematic morphology of massive galaxies in TNG100-1, while the number of major mergers, especially the number of major mergers at $z<1$, do have statistical impact on the morphology of massive galaxies. 

Although massive disc galaxies experienced fewer major mergers than ellipticals, we should note that it does not mean that they do not experience major mergers at all. Since $z=2$, on average a massive disc galaxy selected today experienced nearly one major mergers, and at $z<1$, the average number of major merger is 0.41. More specifically, after $z=1$, 25 of the 83 massive disc galaxies experienced one major merger, 3 experienced two, 1 experienced three, and the rest experienced no major merger. While the galaxies without recent major mergers can be understood to remain a disc morphology, it is interesting to figure out why major mergers, and in some cases even repeat major mergers do not change morphology of the galaxies from disc to elliptical. We will investigate this in detail in Section \ref{sec:MorphologyChange}.

\begin{figure*}
    \hspace{-0.4cm}
    \resizebox{17cm}{!}{\includegraphics{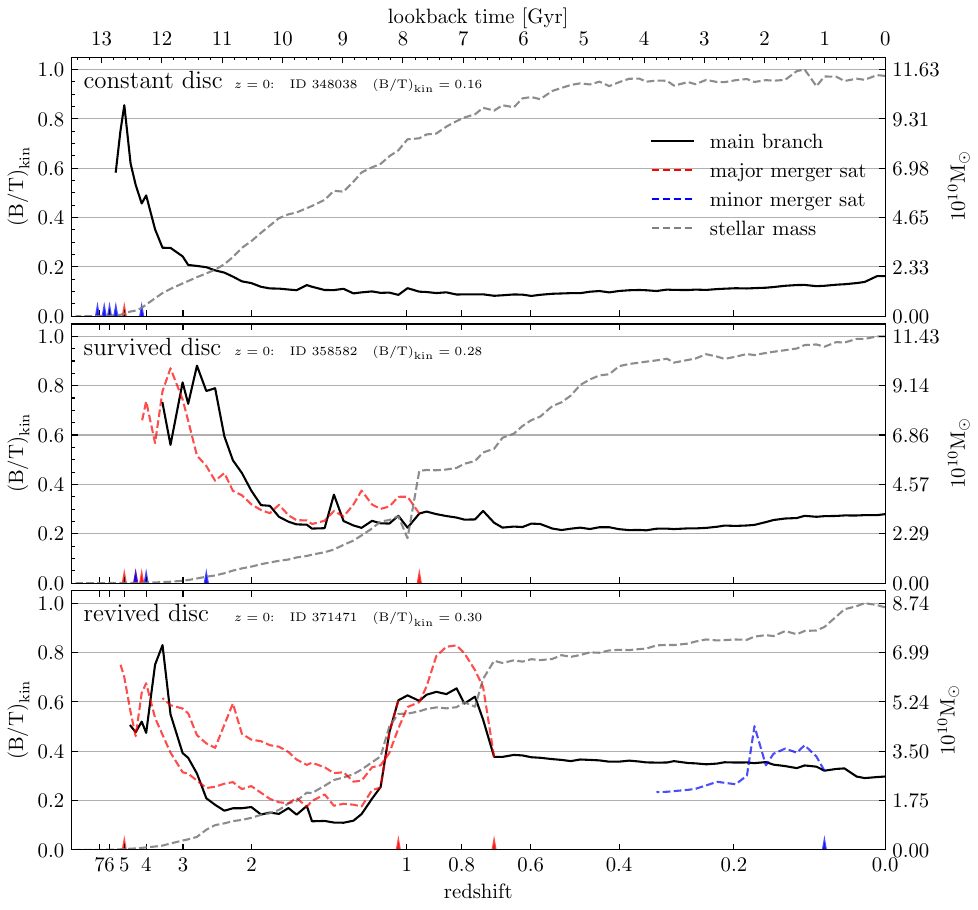}}
        \caption{Morphology evolution and mass assembly history for three example massive disc galaxies with different evolution pathways (from top to bottom: \cd, \sd\, and \rd). The black solid line in each panel shows the \BtoTkin evolution of each galaxy along its main progenitor branch, while gray dashed lines represent the evolution of stellar mass. Mergers happening on the main progenitor branch are marked in each panel. Red/blue triangle symbols on the x-axis mark the time of major/minor mergers. Correspondingly, the red/blue dashed lines indicate the morphological evolution of the satellites before the major/minor merger happen. In the upper left corner of each panel, the ID of the example galaxy in the simulation and \BtoTkin at $z=0$ for each massive disc galaxy are shown.}
        \label{fig:evolution_of_three_typical}
\end{figure*}

\subsection{Different evolution pathways of massive disc galaxies}
\label{subsec:DifferentPathways}

As shown in Fig.~\ref{fig:massive_galaxies_BtoT_kin_evolution}, in general massive disc galaxies maintained a discy morphology after $z\sim2$, with a median \BtoTkin lower than $0.3$. Nevertheless, individual galaxy has various history of morphology evolution, which form the scatters around the median. Especially, it can be seen from the blue shadow area in the Fig.~\ref{fig:massive_galaxies_BtoT_kin_evolution} that, at $z\sim 1$, the progenitors of some massive disc galaxies have \BtoTkin greater than 0.5. We have looked at the \BtoTkin evolution and merger history of each massive disc galaxy in our sample, and find that they do have distinct evolution history in morphology. Galaxies without any major or minor merger after $z\sim 2$ evolve into and remain in a disc state till present day. For the galaxies that have undergone major or minor mergers, some of them also remain in a disc state without significant increase in \BtoTkin till present day, while others have a period of obvious increase in the bulge fraction, but change to a disc galaxy again till today.

To quantitatively classify the different evolution pathways of each massive disc galaxy, we analyse its merger history and \BtoTkin evolution starting from the time when the galaxy first form a kinematic disc-dominant morphology in the simulation. Specifically, this time is defined as the first time when \BtoTkin of the main progenitor drops to be below 0.4 for at least three snapshots\footnote{This time is around $z = 2 - 3$ for most of the massive discs. If this time is later than redshift 1, then we start from $z=1$.}. The evolution pathways afterwards are then classified into three types:

\begin{itemize}
\item[(1)] \cd\,-- the main progenitor branch of the galaxy stays constantly in a disc shape, having \BtoTkin always less than 0.4 and experience no major or minor merger(s);
\item[(2)] \sd\,-- the main progenitors experience major or minor merger(s), but survives as a disc, with \BtoTkin stays below 0.4;
\item[(3)] \rd\,-- \BtoTkin of the main progenitor branch once increases above 0.4, but later drops again to be below 0.4.
\end{itemize}

According to this classification scheme, 7 of our 83 massive disc galaxies are \cds, 31 are \sds, and the remaining 45 are \rds. 
We show in Fig.~\ref{fig:evolution_of_three_typical} one example for each of the above type of massive disc galaxies. In each panel, solid line gives the \BtoTkin evolution of the main progenitors. Grey dashed lines show the evolution of the stellar mass. Major and minor merger events are marked in the panel, with red (major merger) and blue (minor merger) triangles on the x-axis indicating the merger times, and red and blue dashed lines indicating the \BtoTkin evolution of the satellite galaxy before each merger.

In the bottom panel of Fig.~\ref{fig:evolution_of_three_typical}, after the first assembly of a disc galaxy, this \rd\, galaxy have a major merger at $z=1.04$, accompanied with a significant stellar mass growth and an increase of \BtoTkin. Then at $z=0.70$, it experiences another major merger, with another sudden increase in stellar mass. This time after the major merger, however, the galaxy decreases in \BtoTkin, from $> 0.6$ to $< 0.4$, and changes from a bulge-dominant galaxy to a disc-dominant one. 
In the middle panel, the \sd\, galaxy undergoes a major merger at $z=0.95$ and so its stellar mass increases rapidly. However, the kinematic morphology of this galaxy does not change much, with \BtoTkin always being around $0.3$ till today. 
In the top panel, the \cd\, galaxy has a quiet merger history and a relatively smooth stellar mass growth curve. 

For these three different types of massive disc galaxies, their average number of major and minor mergers experienced are also listed in Table~\ref{tab:merger_number}. By definition, \cd\, galaxies have few mergers, especially at low redshifts. \Rd\, galaxies experienced on average both more major and more minor mergers than \sd\, galaxies. 
By looking at the numbers of galaxies for each type of evolution pathway, we see that only 8.4 percent of the kinematically selected massive disc galaxies today have disc morphology because they have quiet merger histories. The large fraction of massive disc galaxies do experience mergers and even major mergers, but stay in disc shape (37.3 percent), or become discs again after a period of being bulge-dominant in history (54.2 percent).

From the numbers presented in Table~\ref{tab:merger_number}, and as can be seen from the example galaxies shown in Fig.~\ref{fig:evolution_of_three_typical}, major mergers can indeed make galaxies elliptical as expected, but do not always turn discs into ellipticals. Some major mergers do not destroy the pre-existing disc structures, and some even transform a bulge-dominant galaxy to a disc one. 
In the next section, we will study why major mergers sometimes change morphology of galaxies dramatically but sometimes do not.

\section{Dependence of morphology change on properties of major mergers}
\label{sec:MorphologyChange}

While major mergers do not always change disc galaxies to ellipticals, previous studies showed that gas-rich mergers or mergers with prograde orbital configuration tend to form disc morphology more frequently, as introduced in the Section \ref{sec:Intro}. When analysing the merger histories of galaxies and by looking at the detailed merger processes, we find that the orbit type of whether the galaxy pair have head-on collision or gradual spiral-in may also have impact on the final morphology of the merger remnant. Therefore in this section, we first study the dependence of morphology change in major mergers on cold gas fraction and orbital configuration, then we focus on and define quantitatively the orbit type of mergers, and study the dependence of morphology change in major mergers on orbit type.

\subsection{Morphology change of the latest major mergers}

As seen in Table~\ref{tab:merger_number}, the numbers of major mergers between massive disc and massive elliptical galaxies differ mostly at $z<1$. These recent major mergers are supposed to relate strongly with the morphology of galaxies at $z=0$. We therefore focus on the latest (and at $z<1$) major merger of each massive galaxy, and study the morphology change therein. 

For the 567 massive galaxies selected in the simulation, 347 galaxies experience at least one major merger after redshift 1, which constitutes the sample we study in this section. For each major merger, we quantify the morphological change by the difference between the kinematic bulge-to-total ratio at 1 Gyr before and at 1 Gyr after the time when two galaxies collide\footnote{We choose the merger time to be the last snapshot at which the two merging galaxies can still be identified separately.} in the simulation. This 2 Gyr time-scale for major merger event is chosen to try to avoid the irregular interaction stages in mergers, and is consistent with previous studies \citep[e.g.][]{jiang2008fitting, kaviraj2011coincidence, martin2018role}.

\begin{figure*}
    \hspace{-0.4cm}
    \resizebox{17cm}{!}{\includegraphics{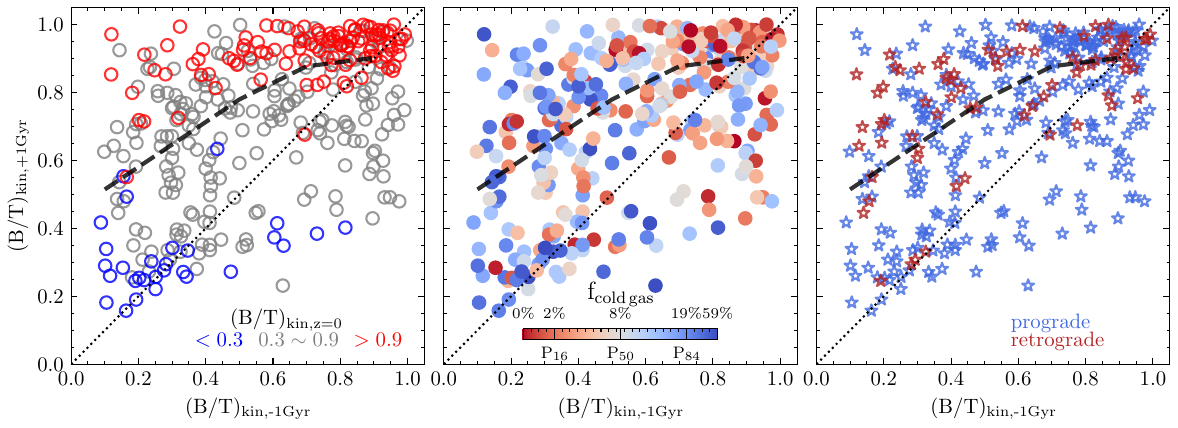}}
        \caption{Morphology change of the latest major mergers for massive galaxies today. In each panel, 347 major mergers are plotted in the plane of kinematic bulge-to-total stellar mass ratio at 1 Gyr before and at 1 Gyr after the merger. Black dashed lines show the median relation. Black dotted lines give a reference for no change of morphology in merger, above which galaxies became more bulge-dominant, and below which galaxies become more disc-dominant after major merger. In the left-hand panel, blue/red circles correspond to the massive disc/elliptical galaxies at $z=0$, and grey circles are for galaxies with intermediate morphology. The middle panel shows the dependence of morphology change on the cold gas fraction of each major merger. The colour of each symbol represents the cold gas fraction as indicated by the colour bar, with more gas-rich mergers being bluer and more gas-poor mergers being redder. In the right-hand panel, prograde and retrograde mergers are indicated by red and blue symbols respectively.}
        \label{fig:BtoT_kin_1GyrBefore_1GyrAfter}
\end{figure*}

In the left-hand panel of Fig.~\ref{fig:BtoT_kin_1GyrBefore_1GyrAfter}, we plot the morphology proxy before and after merger, i.e., \BtoTkinOneGyrBefore and \BtoTkinOneGyrAfter, for the 347 latest major mergers investigated. Blue circles are for the latest major mergers that happen on massive disc galaxies selected at $z=0$. Red circles are for the results of massive elliptical galaxies, and gray circles are for the rest of the massive galaxies. From the figure we see that, in general, major mergers turn galaxies into a more bulge-dominant morphology, where most of the points lie above the black dotted line, as expected. However, there are still a fraction of galaxies decrease in \BtoTkin after major mergers. For massive disc galaxies today, their latest major mergers do not produce bulge-dominant remnants as for massive elliptical galaxies. Some major mergers lead to obvious decrease in bulge component, as we have already seen in examples shown in Fig.~\ref{fig:evolution_of_three_typical}. In the following, we will investigate the difference between major mergers that happen on massive disc galaxies and those on massive ellipticals, to understand further why massive disc galaxies exist at present day.

\subsection{Dependence on cold gas fraction and orbital configuration}

In the IllustrisTNG simulation, each gas particle includes cold/neutral phase and hot/ionized phase.
For a gas particle with star formation rate greater than 0, the cold-phase mass fraction is generally greater than $90\%$ \citep[][]{springel2003cosmological}, and the cold gas mass can be represented by the mass of the star-forming particles \citep[e.g.][]{diemer2018modeling}. 
For gas particles with no star formation, they also contain neutral gas. The neutral gas fraction is provided in TNG100-1, but only for 20 'full' snapshots, which makes it difficult to calculate cold gas mass in these non-star-forming particles for all our sample galaxies. Nevertheless, for the galaxies that are in the 'full' snapshots of TNG100-1, we have checked that whether including or not the cold gas mass in the non-star-forming particles does not affect our derived results. 
Therefore, in the following we take the sum of the star-forming gas particles within a galaxy to represent its cold gas mass.

We first check the dependence of morphology change on cold gas fraction. 
The cold gas fraction of a merger is defined as the ratio between cold gas mass and the sum of cold gas mass and stellar mass, taking into account both merging galaxies, at the time 1 Gyr before the merger happens.

In the middle panel of Fig.~\ref{fig:BtoT_kin_1GyrBefore_1GyrAfter}, we plot again morphology change of the latest major mergers for the massive galaxies selected, with colours from red to blue indicating cold gas fraction of the merging system from gas-poor to gas-rich. For the 347 major mergers we investigate, the median cold gas fraction is $8 \%$, while the 16 and 84 percentiles of the cold gas fraction distribution are $2 \%$ and $19 \%$, as indicated in the colour bar of the middle panel of Fig.~\ref{fig:BtoT_kin_1GyrBefore_1GyrAfter}.

From the figure we see that, most of the elliptical remnants with \BtoTkinOneGyrAfter $>0.9$ are the results of gas-poor mergers, when the progenitor galaxies are bulge-dominant. Remnants with smallest bulges are mostly from gas-rich mergers. Nevertheless, some major mergers with f$_\text{cold\,gas} < 2\%$ produce a \BtoTkin $< 0.4$ remnant like their gas-rich counterparts. 
If we look at the morphology change, indicated by the deviation from the dotted line, the dependence on cold gas fraction seems to be weak. Gas-rich mergers can increase bulge fraction by quite amount, and gas-poor mergers can make bulge-dominant galaxies more discy. 
Therefore in general, morphology change has weak dependence on cold gas fraction, which alone cannot explain why galaxies became a disc after a major merger.

Previous works found that in other hydrodynamical simulations, gas-rich major mergers, even 1:1 mass-ratio mergers, can yield disc-dominant remnants \citep[e.g.][]{hopkins2009disks}. 
\citet{martin2018role} found that when the main progenitor is a disc, merger is typically spinning-down the galaxy regardless of the gas fraction. Such cases are also found in our samples. However, some of our results are not fully consistent with previous works \citep[e.g.][]{peschken2020disc, jackson2020extremely}.  
For example, \citet{jackson2020extremely} claimed that apart from the 30 per cent galaxies which have quiet merger histories, the rest of their massive disc galaxies become discs through a latest merger that happens between a massive spheroidal galaxy and a gas-rich satellite galaxy.
Differences between this work and some previous studies may be the consequences of different subgrid physics models adopted in the simulations, different stellar mass ranges studied, and/or diverse methods used for calculating gas fraction.

In the right-hand panel of Fig.~\ref{fig:BtoT_kin_1GyrBefore_1GyrAfter}, we investigate the effect of orbital configuration of major merger on remnant morphology. The orbital configuration describes whether the angular momentum of the satellite galaxy and the merging orbit in sum are in the same direction as the rotation of the merging central galaxy (prograde) or not (retrograde). Specifically, following the definition proposed by \citet{martin2018role}, we calculate the external angular momentum of a merger by:

\newcommand{\Lext}{{\text{L}}_{\text{external}}}
\newcommand{\Lmain}{{\textbf{L}}_{\text{main}}}
\newcommand{\Lsat}{{\textbf{L}}_{\text{sat}}}
\newcommand{\Lorb}{{\textbf{L}}_{\text{orb}}}

\begin{equation}
    \Lext = |\Lsat|\cos{({\theta}_{\Lmain,\Lsat})} + |\Lorb|\cos{({\theta}_{\Lmain,\Lorb})},
	\label{eq:L_ext}
\end{equation}
where $\Lmain$ and $\Lsat$ are the stellar angular momentum of main progenitor galaxy and satellite galaxy in a merger respectively. The merging orbital angular momentum $\Lorb = \text{M}_\text{sat}(\textbf{r}\times\textbf{v})$ is calculated by the orbit of the satellite relative to the main progenitor. The angle between the angular momentum of main progenitor and satellite (of main progenitor and orbit) is  ${\theta}_{\Lmain,\Lsat}$ (${\theta}_{\Lmain,\Lorb}$).

If $\Lext > 0$ at 1 Gyr before coalescence happens, the major merger is defined as a prograde merger. In contrast, retrograde merger has $\Lext < 0$. We mark in the right-hand panel of Fig.~\ref{fig:BtoT_kin_1GyrBefore_1GyrAfter}, the type of orbital configuration for each major merger we study. As shown, most of the major mergers are prograde (79.0 percent), which can increase and can also decrease the bulge component of merged galaxies. On the other hand, retrograde mergers always make galaxies more bulge-dominant or keep similar morphology as before merger.
Previous works \citep[e.g.][]{hopkins2009disks, martin2018role} suggest that disc structure is more likely to survive after a prograde merger compared to a retrograde one, which is consistent with the trend shown in the right-hand panel of Fig.~\ref{fig:BtoT_kin_1GyrBefore_1GyrAfter}. However, the wide distribution of prograde mergers on the plane of morphology change indicates that the type of orbital configuration of major merger is not able to determine the disc morphology of major merger remnant either. 

\subsection{A strong dependence of remnant morphology on orbit type}
\label{subsec:OrbitType}

While the previous subsection shows that morphology change of major mergers do not depend strongly on cold gas fraction or orbital configuration of the merging system, we study further whether the detailed orbit type of the merger correlates with the morphology change of galaxies, as introduced in Section \ref{sec:Intro}. 
For each of the 347 major mergers selected, we look at in detail the orbit of the satellite galaxy around the merging central galaxy.
We found that for the latest major mergers of massive disc galaxies today, their merger orbits are mostly spiral-in. On the other hand, for massive elliptical galaxies today, their latest major mergers are largely head-on collisions. In the following, we study if there exists quantitative correlation between the type of merger orbit and the morphology change in mergers.

\begin{figure*}
    \animategraphics[width=0.8\linewidth, autoplay, loop, controls]{1}{figures/animation/}{00}{12}
    \caption{Two examples of the orbit of major mergers in our sample. The upper row is for a head-on merger orbit, and the lower row is for a spiral-in merger orbit. Left-hand panels show merging orbit, and right-hand panels give the stellar mass density projection of the merging galaxies. Each frame in this animation corresponds to a snapshot in the simulation, and the time of the snapshot relative to merger time is indicated in the upper right corner of each row. In the left-hand panel of each row, the fixed red circle in the centre indicates the position of the merging central galaxy, and the moving red circle represents the position of the satellite galaxy. At each snapshot, the black arrow shows the velocity of the satellite relative to the central, and the black dashed line marks the relative position. The angle between the black arrow and black line at each snapshot is recorded as $\theta_{1}$, $\theta_{2}$ and so on. The initial angle $\theta_{1}$ and the average angle $\overline{\theta}$ ($= \frac{1}{n} \sum_{i=1}^{n}{\theta_{i}}$) for each orbit, as listed in the upper left corner of each panel in left columns, are used to quantitatively measure the orbit type of major merger in this work. }
    \label{fig:different_orbits}
\end{figure*}

For each major merger, we focus on the merger orbit from 1 Gyr before merger till the time merger finally occurs. In Fig.~\ref{fig:different_orbits}, we show two examples of the merger orbits studied. The upper row gives an example of a head-on merger orbit and the lower row is for a spiral-in merger orbit. At each snapshot of the simulation output, we use the  angle $\theta$ (acute angle) between the position-vector and the velocity-vector of the satellite galaxy relative to the central galaxy, to quantify the orbital direction of the satellite galaxy, as shown in the left-hand panels of Fig.~\ref{fig:different_orbits}. If $\theta={0}^{\circ}$, the satellite is heading to the central straightly, and for $\theta={90}^{\circ}$, the satellite galaxy is on a circular orbit around the central at this point. Then, 
for the merging process from 1 Gyr before merger till the merger finally occurs, we calculate the average $\theta$ to represent the overall orbit type:
\begin{equation}
    \overline{\theta} = \frac{1}{n} \sum_{i=1}^{n}{\theta_{i}}
\end{equation}
where $\theta_{i}$ represents the angle at each snapshot before merger. $\theta_{1}={\theta}_{\text{-1Gyr}}$, is at 1Gyr before merger, and $\theta_{n}$ is at the last snapshot before the two galaxies finally merge into one.

For the orbit shown in the upper panels of Fig.~\ref{fig:different_orbits}, which is close to a head-on collision, ${\theta}_{\text{-1Gyr}}={26.09}^{\circ}$ and $\overline{\theta}={14.33}^{\circ}$. For the case shown in the lower panels which is a spiral-in orbit, ${\theta}_{\text{-1Gyr}}={75.43}^{\circ}$ and $\overline{\theta}={81.15}^{\circ}$.
While the satellite galaxy may not yet be in its stable merging orbit at a given snapshot before merger, and the merger orbit of the satellite studied may be disturbed at certain snapshots by other smaller satellites around the same central galaxy, we believe the average angle $\overline{\theta}$ represents better the overall orbit type than the angle at a certain time, for example ${\theta}_{\text{-1Gyr}}$ (which is confirmed in Fig.~\ref{fig:BtoT_kin_versus_theta}). Therefore we focus on $\overline{\theta}$ to study the dependence of  morphological change on orbit type.

\begin{figure}
	\includegraphics[width=\columnwidth]{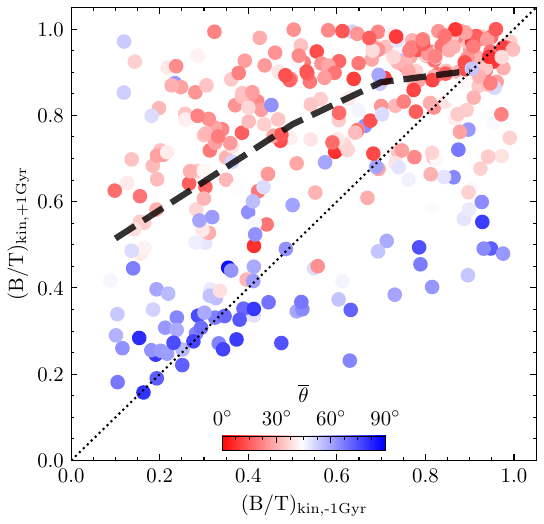}
    \caption{Similar as Fig.~\ref{fig:BtoT_kin_1GyrBefore_1GyrAfter}, morphology change of the latest major mergers of massive galaxies today. Symbols are colour coded with the orbit type indicated by the average angle $\overline{\theta}$ (see Section \ref{subsec:OrbitType} for the detailed definition). }
    \label{fig:BtoT1GyrBefore_vs_BtoT1GyrAfter_vs_mean_theta}
\end{figure}

\begin{figure*}
    \hspace{-0.4cm}
    \resizebox{17cm}{!}{\includegraphics{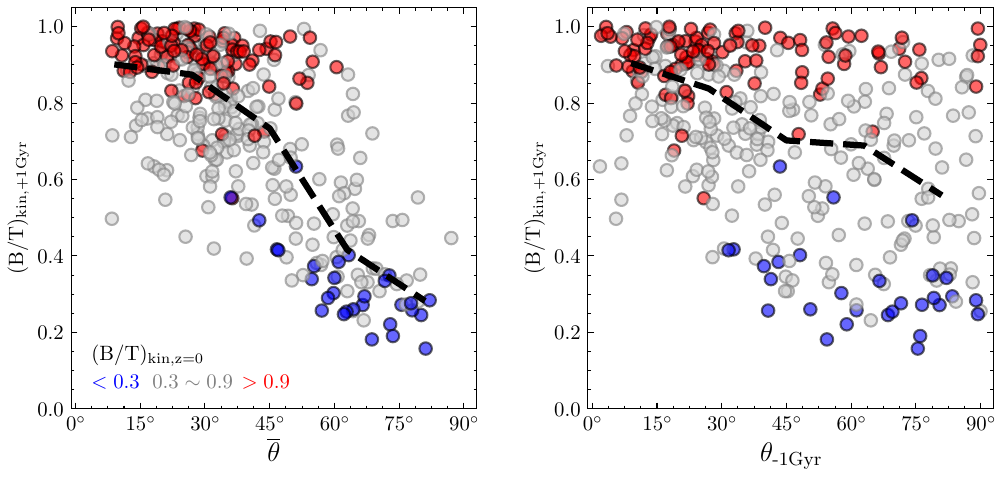}}
    \caption{The remnant morphology after major merger as a function of $\overline{\theta}$ (left-hand panel) and ${\theta}_{\text{-1Gyr}}$ (right-hand panel). In each panel, the median of the relation is indicated by black dashed line. The blue and red points show the major mergers corresponding to the massive disc and elliptical galaxies selected at $z=0$ respectively. Major mergers of other galaxies are shown by gray points.}
    \label{fig:BtoT_kin_versus_theta}
\end{figure*}

In Fig.~\ref{fig:BtoT1GyrBefore_vs_BtoT1GyrAfter_vs_mean_theta}, morphology change during major mergers are plotted again, similar as shown in Fig.~\ref{fig:BtoT_kin_1GyrBefore_1GyrAfter}, but this time colour coded with the average orbit angle $\overline{\theta}$, with bluer symbols representing more circular orbit and redder symbols representing more head-on orbit. The results show a clear dependence of the morphology of merger remnant on orbit type. 
Almost all major mergers with a remnant of \BtoTkinOneGyrAfter $< 0.4$ correspond to merger orbits with $\overline{\theta} > \sim{50}^{\circ}$, no matter what morphology the initial galaxy is before merger. On the other hand, most bulge-dominated remnants are results of major mergers of small $\overline{\theta}$, having (nearly) head-on collisions. If we look at the morphology change during these mergers, i.e., the deviation away from the dotted line in Fig.~\ref{fig:BtoT1GyrBefore_vs_BtoT1GyrAfter_vs_mean_theta}, prominent bulge increase mostly happens in head-on mergers with low $\overline{\theta}$, and obvious disc increase or bulge decrease mostly happens in spiral-in mergers with large $\overline{\theta}$. Note that the relative large number of galaxies concentrated in the left lower and right upper corners of the figure is quite possibly due to the limited range of \BtoTkin.

Fig.~\ref{fig:BtoT1GyrBefore_vs_BtoT1GyrAfter_vs_mean_theta} shows that the morphology of remnant after major merger is strongly correlated with the type of merger orbit, much more than the correlation with either cold gas fraction or orbital configuration of the system as shown in Fig.~\ref{fig:BtoT_kin_1GyrBefore_1GyrAfter}. In the left-hand panel of Fig.~\ref{fig:BtoT_kin_versus_theta}, we plot the correlation between $\overline{\theta}$ and \BtoTkinOneGyrAfter directly. The median of this relation is indicated by the dashed black line, which clearly shows the strong dependence of morphology of merger remnant on orbit type. The mergers that belong to the massive disc galaxies at $z=0$ are plotted in blue, most of which have large $\overline{\theta}$ and are spiral-in merger orbits. Mergers of massive elliptical galaxies are plotted in red, most of which have orbits close to head-on. Although the morphology of galaxies can still change by some amount after the latest major merger, the left-hand panel of Fig.~\ref{fig:BtoT_kin_versus_theta} shows that massive disc galaxies that have experienced recent major mergers can survive as disc morphology mostly because the orbit of its latest major merger is a spiral-in type. 

In the right-hand panel of Fig.~\ref{fig:BtoT_kin_versus_theta}, we check the relation between \BtoTkinOneGyrAfter and the orbit angle at 1 Gyr before merger ${\theta}_{\text{-1Gyr}}$. As shown, the correlation still exists, but is weaker and with larger scatter compared to that of the left-hand panel, indicating that the average orbit angle $\overline{\theta}$ is a better indicator of orbit type than ${\theta}_{\text{-1Gyr}}$ to be correlated with morphology of merger remnant.

\section{Discussion and Conclusions}
\label{sec:Conclusions}

In this work, we study the formation history of massive disc galaxies in the hydrodynamical simulation TNG100-1. We select massive galaxies with stellar mass greater than $8\times10^{10}$ \Msun at $z=0$, and use the kinematic bulge-to-total stellar mass ratio \BtoTkin as an indicator to quantify galaxy morphology. Massive galaxies with \BtoTkin $< 0.3$ are selected as our massive disc galaxies sample, and those with \BtoTkin $> 0.9$ are selected as massive elliptical galaxies sample for comparison. We study the morphology evolution and merger history of different types of massive galaxies, and investigate the relation between various properties of major mergers with the morphology change during these mergers, to try to understand why some galaxies experience recent major mergers but stay in a disc-dominant morphology. Our main results are as follows.

\begin{itemize}
\item[(i)] On average, the morphological evolution histories of massive disc and massive elliptical galaxies today start to deviate at $z\sim2$. For their main progenitors at $z\sim2$, the median \BtoTkin is around 0.3 for all types of galaxies. Then the main progenitors of massive disc galaxies maintained a discy morphology, while those of massive ellipticals increase dramatically in \BtoTkin till today (Fig.\ref{fig:massive_galaxies_BtoT_kin_evolution}).

\item[(ii)] Massive disc and massive elliptical galaxies have similar minor merger histories, with about the same number of minor mergers at different redshift intervals. For major merger history, massive disc galaxies experienced $\sim 1$ less major merger than their elliptical counterparts, mostly contributed by major merger at $z<1$ (Table~\ref{tab:merger_number}).

\item[(iii)] Based on the morphology evolution and merger history after the time of the first assembly of a disc-like structure, the 83 massive disc galaxies we select can be divided into three types. 7 of them (8.4\%) have quiet merger history and keep having small \BtoTkin. 45 of them (54.2\%) have obvious increase in bulge-fraction in history, but eventually regain a disc structure as seen today. The remaining 31 (37.3\%) experience recent minor and/or major mergers, but preserve their pre-existing disc morphology until today (Fig.~\ref{fig:evolution_of_three_typical}).

\item[(iv)] For the latest major mergers of the massive galaxies, morphology change during major mergers is found to have small dependence on cold gas fraction and orbital configuration of the merging system (Fig.~\ref{fig:BtoT_kin_1GyrBefore_1GyrAfter}). There exists, however, a strong dependence of morphology of merger remnant on orbit type, in the sense that the merging orbit is a head-on collision or a spiral-in falling (Fig.~\ref{fig:BtoT1GyrBefore_vs_BtoT1GyrAfter_vs_mean_theta}). In general, head-on collisions lead to bulge-dominant merger remnants and gradual spiral-in orbits lead to disc-dominated remnants. This correlation is stronger when using the average of the orbit angle from 1 Gyr before merger till merger time ($\overline{\theta}$), than using the orbit angle at a given specific time (e.g. ${\theta}_{\text{-1Gyr}}$), as shown in Fig.~\ref{fig:BtoT_kin_versus_theta}.
\end{itemize}

The findings above answer the question why massive disc galaxies exist today. For TNG100-1 massive discs, first, a small fraction of them keep disc morphology because they experience few mergers in history. Then, for the rest of the galaxies that have recent mergers, especially major mergers, they have disc morphology mainly because their latest major mergers have the orbit type of gradual spiral-in, which normally lead to disc-dominant merger remnants regardless of the original morphology of galaxies before merger.

Our results show that the morphology of merger remnant has a much stronger dependence on the type of merger orbit than cold gas fraction or orbital configuration. Nevertheless, we should note that there is still quite a scatter in the relation between orbit type indicator $\overline{\theta}$ and \BtoTkinOneGyrAfter as seen in Fig.~\ref{fig:BtoT_kin_versus_theta}. Also, in the upper left corner of Fig.~\ref{fig:BtoT1GyrBefore_vs_BtoT1GyrAfter_vs_mean_theta}, we can see that some spiral-in major mergers of relative large $\overline{\theta}$ also boost the bulge fraction of galaxies. Therefore, to fully explore the relation between properties of major mergers and the morphology of merger remnant, we need to take into account the effect of orbit type, mass ratio, cold gas fraction, orbital configuration, and other possible potential effects all together, which will be investigated in a future work.

While this work has been focusing on massive disc galaxies at $z=0$, massive disc galaxies are also observed at higher redshift \citep[e.g.][]{xu2020star, xu2021giant}. Especially, \citet{xu2021giant} found a massive red disc galaxy at $z=0.76$, and failed to find any simulation counterpart in the Illustris-1 and the TNG100-1. Future studies should be performed to understand the formation of massive disc galaxies at higher redshifts, and to explain in particular the quenching of massive disc galaxy at different redshifts.


\section*{Acknowledgements}

We thank the anonymous referee for a prompt and constructive report. 
We thank Quan Guo for help in downloading data from the IllustrisTNG database, and thank Dylan Nelson and Yingjie Jing for helping us to use the IllustrisTNG data.
We also thank Shi Shao, Yi He, Haonan Zheng and Jianhui Qiu for their helpful discussions and comments.
This work was supported by the National Natural Science Foundation of China (NSFC) under grant 11988101, and the National Key Program for Science and Technology Research Development of China 2018YFE0202902.
The IllustrisTNG simulations were undertaken with compute time awarded by the Gauss Centre for Supercomputing (GCS) under GCS Large-Scale Projects GCS-ILLU and GCS-DWAR on the GCS share of the supercomputer Hazel Hen at the High Performance Computing Center Stuttgart (HLRS), as well as on the machines of the Max Planck Computing and Data Facility (MPCDF) in Garching, Germany.

\section*{Data Availability}

The IllustrisTNG simulations, including the TNG100-1 used in this article, are publicly available and accessible at \url{https://www.tng-project.org/data/}. The data produced in this article will be shared on reasonable request to the corresponding author.



\bibliographystyle{mnras}
\bibliography{references} 

\begin{thebibliography}{}
\makeatletter
\relax
\def\mn@urlcharsother{\let\do\@makeother \do\$\do\&\do\#\do\^\do\_\do\%\do\~}
\def\mn@doi{\begingroup\mn@urlcharsother \@ifnextchar [ {\mn@doi@}
  {\mn@doi@[]}}
\def\mn@doi@[#1]#2{\def\@tempa{#1}\ifx\@tempa\@empty \href
  {http://dx.doi.org/#2} {doi:#2}\else \href {http://dx.doi.org/#2} {#1}\fi
  \endgroup}
\def\mn@eprint#1#2{\mn@eprint@#1:#2::\@nil}
\def\mn@eprint@arXiv#1{\href {http://arxiv.org/abs/#1} {{\tt arXiv:#1}}}
\def\mn@eprint@dblp#1{\href {http://dblp.uni-trier.de/rec/bibtex/#1.xml}
  {dblp:#1}}
\def\mn@eprint@#1:#2:#3:#4\@nil{\def\@tempa {#1}\def\@tempb {#2}\def\@tempc
  {#3}\ifx \@tempc \@empty \let \@tempc \@tempb \let \@tempb \@tempa \fi \ifx
  \@tempb \@empty \def\@tempb {arXiv}\fi \@ifundefined
  {mn@eprint@\@tempb}{\@tempb:\@tempc}{\expandafter \expandafter \csname
  mn@eprint@\@tempb\endcsname \expandafter{\@tempc}}}

\bibitem[\protect\citeauthoryear{Athanassoula, Rodionov, Peschken  \&
  Lambert}{Athanassoula et~al.}{2016}]{athanassoula2016forming}
Athanassoula E.,  Rodionov S.,  Peschken N.,   Lambert J.,  2016, The
  Astrophysical Journal, 821, 90

\bibitem[\protect\citeauthoryear{Barnes \& Hernquist}{Barnes \&
  Hernquist}{1996}]{barnes1996transformations}
Barnes J.~E.,  Hernquist L.,  1996, The Astrophysical Journal, 471, 115

\bibitem[\protect\citeauthoryear{Buitrago, Trujillo, Conselice  \&
  H{\"a}u{\ss}ler}{Buitrago et~al.}{2013}]{buitrago2013early}
Buitrago F.,  Trujillo I.,  Conselice C.~J.,   H{\"a}u{\ss}ler B.,  2013,
  Monthly Notices of the Royal Astronomical Society, 428, 1460

\bibitem[\protect\citeauthoryear{Conselice}{Conselice}{2014}]{conselice2014evolution}
Conselice C.~J.,  2014, Annual Review of Astronomy and Astrophysics, 52, 291

\bibitem[\protect\citeauthoryear{Correa \& Schaye}{Correa \&
  Schaye}{2020}]{correa2020dependence}
Correa C.~A.,  Schaye J.,  2020, Monthly Notices of the Royal Astronomical
  Society, 499, 3578

\bibitem[\protect\citeauthoryear{Cox, Dutta, Di~Matteo, Hernquist, Hopkins,
  Robertson  \& Springel}{Cox et~al.}{2006}]{cox2006kinematic}
Cox T.,  Dutta S.~N.,  Di~Matteo T.,  Hernquist L.,  Hopkins P.~F.,  Robertson
  B.,   Springel V.,  2006, The Astrophysical Journal, 650, 791

\bibitem[\protect\citeauthoryear{Diemer et~al.,}{Diemer
  et~al.}{2018}]{diemer2018modeling}
Diemer B.,  et~al., 2018, The Astrophysical Journal Supplement Series, 238, 33

\bibitem[\protect\citeauthoryear{Dolag, Borgani, Murante  \& Springel}{Dolag
  et~al.}{2009}]{dolag2009substructures}
Dolag K.,  Borgani S.,  Murante G.,   Springel V.,  2009, Monthly Notices of
  the Royal Astronomical Society, 399, 497

\bibitem[\protect\citeauthoryear{Doroshkevich}{Doroshkevich}{1970}]{doroshkevich1970spatial}
Doroshkevich A.,  1970, Astrophysics, 6, 320

\bibitem[\protect\citeauthoryear{Eliche-Moral, Rodr{\'\i}guez-P{\'e}rez,
  Borlaff, Querejeta  \& Tapia}{Eliche-Moral
  et~al.}{2018}]{eliche2018formation}
Eliche-Moral M. d.~C.,  Rodr{\'\i}guez-P{\'e}rez C.,  Borlaff A.,  Querejeta
  M.,   Tapia T.,  2018, Astronomy \& astrophysics, 617, A113

\bibitem[\protect\citeauthoryear{Fall}{Fall}{1979}]{fall1979dissipation}
Fall S.~M.,  1979, Nature, 281, 200

\bibitem[\protect\citeauthoryear{Fall \& Efstathiou}{Fall \&
  Efstathiou}{1980}]{fall1980formation}
Fall S.~M.,  Efstathiou G.,  1980, Monthly Notices of the Royal Astronomical
  Society, 193, 189

\bibitem[\protect\citeauthoryear{Font, McCarthy, Le~Brun, Crain  \&
  Kelvin}{Font et~al.}{2017}]{font2017diversity}
Font A.~S.,  McCarthy I.~G.,  Le~Brun A.~M.,  Crain R.~A.,   Kelvin L.~S.,
  2017, Publications of the Astronomical Society of Australia, 34

\bibitem[\protect\citeauthoryear{Garrison-Kimmel et~al.,}{Garrison-Kimmel
  et~al.}{2018}]{garrison2018origin}
Garrison-Kimmel S.,  et~al., 2018, Monthly Notices of the Royal Astronomical
  Society, 481, 4133

\bibitem[\protect\citeauthoryear{Genel et~al.,}{Genel
  et~al.}{2014}]{genel2014introducing}
Genel S.,  et~al., 2014, Monthly Notices of the Royal Astronomical Society,
  445, 175

\bibitem[\protect\citeauthoryear{Genel, Fall, Hernquist, Vogelsberger, Snyder,
  Rodriguez-Gomez, Sijacki  \& Springel}{Genel
  et~al.}{2015}]{genel2015galactic}
Genel S.,  Fall S.~M.,  Hernquist L.,  Vogelsberger M.,  Snyder G.~F.,
  Rodriguez-Gomez V.,  Sijacki D.,   Springel V.,  2015, The Astrophysical
  Journal Letters, 804, L40

\bibitem[\protect\citeauthoryear{Genel et~al.,}{Genel
  et~al.}{2018}]{genel2018size}
Genel S.,  et~al., 2018, Monthly Notices of the Royal Astronomical Society,
  474, 3976

\bibitem[\protect\citeauthoryear{Governato et~al.,}{Governato
  et~al.}{2009}]{governato2009forming}
Governato F.,  et~al., 2009, Monthly Notices of the Royal Astronomical Society,
  398, 312

\bibitem[\protect\citeauthoryear{Hopkins, Cox, Younger  \& Hernquist}{Hopkins
  et~al.}{2009}]{hopkins2009disks}
Hopkins P.~F.,  Cox T.~J.,  Younger J.~D.,   Hernquist L.,  2009, The
  Astrophysical Journal, 691, 1168

\bibitem[\protect\citeauthoryear{Hubble}{Hubble}{1926}]{hubble1926extragalactic}
Hubble E.~P.,  1926, The Astrophysical Journal, 64

\bibitem[\protect\citeauthoryear{Jackson, Martin, Kaviraj, Laigle, Devriendt,
  Dubois  \& Pichon}{Jackson et~al.}{2020}]{jackson2020extremely}
Jackson R.~A.,  Martin G.,  Kaviraj S.,  Laigle C.,  Devriendt J.,  Dubois Y.,
   Pichon C.,  2020, Monthly Notices of the Royal Astronomical Society, 494,
  5568

\bibitem[\protect\citeauthoryear{Jiang, Jing, Faltenbacher, Lin  \& Li}{Jiang
  et~al.}{2008}]{jiang2008fitting}
Jiang C.,  Jing Y.,  Faltenbacher A.,  Lin W.,   Li C.,  2008, The
  Astrophysical Journal, 675, 1095

\bibitem[\protect\citeauthoryear{Kaviraj, Tan, Ellis  \& Silk}{Kaviraj
  et~al.}{2011}]{kaviraj2011coincidence}
Kaviraj S.,  Tan K.-M.,  Ellis R.~S.,   Silk J.,  2011, Monthly Notices of the
  Royal Astronomical Society, 411, 2148

\bibitem[\protect\citeauthoryear{Lotz, Jonsson, Cox  \& Primack}{Lotz
  et~al.}{2008}]{lotz2008galaxy}
Lotz J.~M.,  Jonsson P.,  Cox T.,   Primack J.~R.,  2008, Monthly Notices of
  the Royal Astronomical Society, 391, 1137

\bibitem[\protect\citeauthoryear{Luo, Li, Kang, Li  \& Wang}{Luo
  et~al.}{2020}]{luo2020has}
Luo Y.,  Li Z.,  Kang X.,  Li Z.,   Wang P.,  2020, Monthly Notices of the
  Royal Astronomical Society: Letters, 496, L116

\bibitem[\protect\citeauthoryear{Marinacci et~al.,}{Marinacci
  et~al.}{2018}]{marinacci2018first}
Marinacci F.,  et~al., 2018, Monthly Notices of the Royal Astronomical Society,
  480, 5113

\bibitem[\protect\citeauthoryear{Martin, Kaviraj, Devriendt, Dubois  \&
  Pichon}{Martin et~al.}{2018}]{martin2018role}
Martin G.,  Kaviraj S.,  Devriendt J.,  Dubois Y.,   Pichon C.,  2018, Monthly
  Notices of the Royal Astronomical Society, 480, 2266

\bibitem[\protect\citeauthoryear{Mo, Mao  \& White}{Mo
  et~al.}{1998}]{mo1998formation}
Mo H.,  Mao S.,   White S.~D.,  1998, Monthly Notices of the Royal Astronomical
  Society, 295, 319

\bibitem[\protect\citeauthoryear{Naab \& Burkert}{Naab \&
  Burkert}{2003}]{naab2003statistical}
Naab T.,  Burkert A.,  2003, The Astrophysical Journal, 597, 893

\bibitem[\protect\citeauthoryear{Naiman et~al.,}{Naiman
  et~al.}{2018}]{naiman2018first}
Naiman J.~P.,  et~al., 2018, Monthly Notices of the Royal Astronomical Society,
  477, 1206

\bibitem[\protect\citeauthoryear{Nelson et~al.,}{Nelson
  et~al.}{2018}]{nelson2018first}
Nelson D.,  et~al., 2018, Monthly Notices of the Royal Astronomical Society,
  475, 624

\bibitem[\protect\citeauthoryear{Nelson et~al.,}{Nelson
  et~al.}{2019}]{nelson2019illustristng}
Nelson D.,  et~al., 2019, Computational Astrophysics and Cosmology, 6, 1

\bibitem[\protect\citeauthoryear{Ogle, Lanz, Nader  \& Helou}{Ogle
  et~al.}{2016}]{ogle2016superluminous}
Ogle P.~M.,  Lanz L.,  Nader C.,   Helou G.,  2016, The Astrophysical Journal,
  817, 109

\bibitem[\protect\citeauthoryear{Ogle, Lanz, Appleton, Helou  \&
  Mazzarella}{Ogle et~al.}{2019}]{ogle2019catalog}
Ogle P.~M.,  Lanz L.,  Appleton P.~N.,  Helou G.,   Mazzarella J.,  2019, The
  Astrophysical Journal Supplement Series, 243, 14

\bibitem[\protect\citeauthoryear{Peebles}{Peebles}{1969}]{peebles1969origin}
Peebles P.,  1969, The Astrophysical Journal, 155, 393

\bibitem[\protect\citeauthoryear{Peschken, {\L}okas  \& Athanassoula}{Peschken
  et~al.}{2020}]{peschken2020disc}
Peschken N.,  {\L}okas E.~L.,   Athanassoula E.,  2020, Monthly Notices of the
  Royal Astronomical Society, 493, 1375

\bibitem[\protect\citeauthoryear{Pillepich et~al.,}{Pillepich
  et~al.}{2018a}]{pillepich2018simulating}
Pillepich A.,  et~al., 2018a, Monthly Notices of the Royal Astronomical
  Society, 473, 4077

\bibitem[\protect\citeauthoryear{Pillepich et~al.,}{Pillepich
  et~al.}{2018b}]{pillepich2018first}
Pillepich A.,  et~al., 2018b, Monthly Notices of the Royal Astronomical
  Society, 475, 648

\bibitem[\protect\citeauthoryear{Robertson, Bullock, Cox, Di~Matteo, Hernquist,
  Springel  \& Yoshida}{Robertson et~al.}{2006}]{robertson2006merger}
Robertson B.,  Bullock J.~S.,  Cox T.~J.,  Di~Matteo T.,  Hernquist L.,
  Springel V.,   Yoshida N.,  2006, The Astrophysical Journal, 645, 986

\bibitem[\protect\citeauthoryear{Rodriguez-Gomez et~al.,}{Rodriguez-Gomez
  et~al.}{2015}]{rodriguez2015merger}
Rodriguez-Gomez V.,  et~al., 2015, Monthly Notices of the Royal Astronomical
  Society, 449, 49

\bibitem[\protect\citeauthoryear{Saburova, Chilingarian, Katkov, Egorov,
  Kasparova, Khoperskov, Uklein  \& Vozyakova}{Saburova
  et~al.}{2018}]{saburova2018malin}
Saburova A.~S.,  Chilingarian I.~V.,  Katkov I.~Y.,  Egorov O.~V.,  Kasparova
  A.~V.,  Khoperskov S.~A.,  Uklein R.~I.,   Vozyakova O.~V.,  2018, Monthly
  Notices of the Royal Astronomical Society, 481, 3534

\bibitem[\protect\citeauthoryear{Saburova, Chilingarian, Kasparova,
  Sil’chenko, Grishin, Katkov  \& Uklein}{Saburova
  et~al.}{2021}]{saburova2021observational}
Saburova A.~S.,  Chilingarian I.~V.,  Kasparova A.~V.,  Sil’chenko O.~K.,
  Grishin K.~A.,  Katkov I.~Y.,   Uklein R.~I.,  2021, Monthly Notices of the
  Royal Astronomical Society, 503, 830

\bibitem[\protect\citeauthoryear{Sijacki, Vogelsberger, Genel, Springel,
  Torrey, Snyder, Nelson  \& Hernquist}{Sijacki
  et~al.}{2015}]{sijacki2015illustris}
Sijacki D.,  Vogelsberger M.,  Genel S.,  Springel V.,  Torrey P.,  Snyder
  G.~F.,  Nelson D.,   Hernquist L.,  2015, Monthly Notices of the Royal
  Astronomical Society, 452, 575

\bibitem[\protect\citeauthoryear{Sparre \& Springel}{Sparre \&
  Springel}{2017}]{sparre2017unorthodox}
Sparre M.,  Springel V.,  2017, Monthly Notices of the Royal Astronomical
  Society, 470, 3946

\bibitem[\protect\citeauthoryear{Springel}{Springel}{2010}]{springel2010pur}
Springel V.,  2010, Monthly Notices of the Royal Astronomical Society, 401, 791

\bibitem[\protect\citeauthoryear{Springel \& Hernquist}{Springel \&
  Hernquist}{2003}]{springel2003cosmological}
Springel V.,  Hernquist L.,  2003, Monthly Notices of the Royal Astronomical
  Society, 339, 289

\bibitem[\protect\citeauthoryear{Springel \& Hernquist}{Springel \&
  Hernquist}{2005}]{springel2005formation}
Springel V.,  Hernquist L.,  2005, The Astrophysical Journal Letters, 622, L9

\bibitem[\protect\citeauthoryear{Springel et~al.,}{Springel
  et~al.}{2005}]{springel2005simulations}
Springel V.,  et~al., 2005, nature, 435, 629

\bibitem[\protect\citeauthoryear{Springel et~al.,}{Springel
  et~al.}{2018}]{springel2018first}
Springel V.,  et~al., 2018, Monthly Notices of the Royal Astronomical Society,
  475, 676

\bibitem[\protect\citeauthoryear{Tacchella et~al.,}{Tacchella
  et~al.}{2019}]{tacchella2019morphology}
Tacchella S.,  et~al., 2019, Monthly Notices of the Royal Astronomical Society,
  487, 5416

\bibitem[\protect\citeauthoryear{Toomre}{Toomre}{1977}]{toomre1977theories}
Toomre A.,  1977, Annual Review of Astronomy and Astrophysics, 15, 437

\bibitem[\protect\citeauthoryear{Vogelsberger et~al.,}{Vogelsberger
  et~al.}{2014a}]{vogelsberger2014introducing}
Vogelsberger M.,  et~al., 2014a, Monthly Notices of the Royal Astronomical
  Society, 444, 1518

\bibitem[\protect\citeauthoryear{Vogelsberger et~al.,}{Vogelsberger
  et~al.}{2014b}]{vogelsberger2014properties}
Vogelsberger M.,  et~al., 2014b, Nature, 509, 177

\bibitem[\protect\citeauthoryear{Wang, Xu, Gao, Guo, Qu  \& Pan}{Wang
  et~al.}{2019}]{wang2019comparing}
Wang L.,  Xu D.,  Gao L.,  Guo Q.,  Qu Y.,   Pan J.,  2019, Monthly Notices of
  the Royal Astronomical Society, 485, 2083

\bibitem[\protect\citeauthoryear{Weinberger et~al.,}{Weinberger
  et~al.}{2016}]{weinberger2016simulating}
Weinberger R.,  et~al., 2016, Monthly Notices of the Royal Astronomical
  Society, 465, 3291

\bibitem[\protect\citeauthoryear{White \& Rees}{White \&
  Rees}{1978}]{white1978core}
White S.~D.,  Rees M.~J.,  1978, Monthly Notices of the Royal Astronomical
  Society, 183, 341

\bibitem[\protect\citeauthoryear{Xu, Liu, Jing, Wang  \& Lu}{Xu
  et~al.}{2020}]{xu2020star}
Xu K.,  Liu C.,  Jing Y.,  Wang Y.,   Lu S.,  2020, The Astrophysical Journal,
  895, 100

\bibitem[\protect\citeauthoryear{Xu, Liu, Jing, Sawicki  \& Gwyn}{Xu
  et~al.}{2021}]{xu2021giant}
Xu K.,  Liu C.,  Jing Y.,  Sawicki M.,   Gwyn S.,  2021, SCIENCE CHINA Physics,
  Mechanics \& Astronomy, 64, 1

\bibitem[\protect\citeauthoryear{van~der Wel et~al.,}{van~der Wel
  et~al.}{2014}]{van20143d}
van~der Wel A.,  et~al., 2014, The Astrophysical Journal, 788, 28

\makeatother
\end{thebibliography}







\bsp	
\label{lastpage}
\end{document}